\documentclass[11pt, a4paper]{article}
\usepackage{jheppub}
\usepackage{bbm}
\usepackage{amsmath}
\usepackage{amssymb}
\usepackage{amsfonts}
\usepackage{mathrsfs}
\usepackage{graphicx}
\usepackage[11pt]{moresize}
\usepackage[normalem]{ulem}
\usepackage[dvipsnames]{xcolor}
\usepackage[]{hyperref}
\usepackage{orcidlink}
\usepackage{relsize}
\usepackage{adjustbox}
\usepackage[verbose]{placeins}
\usepackage{multirow}
\usepackage{float}
\usepackage{footnote}
\usepackage{tablefootnote}
\usepackage{lineno}
\usepackage{makecell}
\usepackage{subfigure}
\hypersetup{
 bookmarks  = true,      % show bookmarks bar?
 unicode    = false,     % non-Latin characters in Acrobat?s bookmarks
 pdfcreator = {RevTeX},  % documentclass
 colorlinks = true,      % false: boxed links; true: colored links
 linkcolor  = red,       % color of internal links
 citecolor  = blue,      % color of links to bibliography
 filecolor  = black,     % color of file links
 urlcolor   = blue,      % color of external links
}
\makeatletter
\renewcommand{\fnum@table}{\textbf{\tablename~\thetable}}
\renewcommand{\fnum@figure}{\textbf{\figurename~\thefigure}}

\makeatother

\newenvironment{conditions*}
{\par\vspace{\abovedisplayskip}\noindent
  \tabularx{\columnwidth}{>{$}l<{$} @{${}={}$} >{\raggedright\arraybackslash}X}}
{\endtabularx\par\vspace{\belowdisplayskip}}
\title{Probing scalar non-standard interaction of supernova neutrinos in next-generation neutrino experiments}

\author[a]{Sudipta Das,}
\author[a]{Mary Hall Reno}

\affiliation[a]{Department of Physics and Astronomy,  University of Iowa, Iowa City, IA 52242, USA}

\emailAdd{sudipta-das@uiowa.edu}
\emailAdd{mary-hall-reno@uiowa.edu}

\abstract
{
A new neutrino-matter interaction can potentially affect neutrino propagation through matter. In this work, we explore the impact of a flavor-conserving scalar-mediated non-standard neutrino interaction in the supernova neutrino flux. We observe that the presence of a scalar interaction involving muon and tau neutrinos (parameterized as $\eta_{\mu\mu}$ and $\eta_{\tau\tau}$, respectively) can invert the neutrino mass eigenstate in which three neutrino flavor states are produced inside the supernova core, resulting in a significant modification of the electron neutrino flux from the supernova reaching the Earth. In the context of the DUNE experiment, we estimate the number of supernova neutrino events in the presence of scalar non-standard neutrino interaction $\eta_{\mu\mu}$ or $\eta_{\tau\tau}$ and contrast with the case without scalar-mediated non-standard interactions. Our results indicate that such scalar interactions introduce a new degeneracy in the measurement of neutrino mass ordering from supernova neutrinos. We show how the $\bar{\nu}_e$ event distribution in Hyper-Kamiokande experiment may help resolve the degeneracy between a model with new scalar interactions for normal ordered neutrino masses and the standard model with inverted mass ordering for a galactic supernova.
}

\begin{document}
\maketitle
%======================
\section{Introduction}
\label{sec:intro}
%================

The detection of neutrino events from the galactic supernova 1987A  in February 1987 opened a new avenue in the field of neutrino astrophysics. Approximately 25 neutrino events were observed at Kamiokande II~\cite{Kamiokande-II:1987idp}, IMB~\cite{PhysRevLett.58.1494}, and Baksan neutrino observatory~\cite{ALEXEYEV1988209}, providing unprecedented insights into the dynamics of the stellar explosion and validating many theoretical predictions~\cite{annurev:/content/journals/10.1146/annurev.ns.40.120190.001145}. Even this small number of detected events unraveled crucial information about  neutrino transport through the supernova core and put stringent constraints on various neutrino properties such as neutrino decay~\cite{Dodelson:1992tv}, neutrino mass~\cite{PhysRevLett.58.1906}, active-sterile neutrino mixing~\cite{Mastrototaro:2019vug}, among others~\cite{Martinez-Soler:2021unz}. Nearly four decades later, our theoretical understanding of the physical processes driving the supernova explosion has been advanced significantly~\cite{Mirizzi:2015eza,Janka:2017vlw,Horiuchi:2018ofe}. Additionally, on the experimental front, a new generation of neutrino detectors with high-precision detection abilities are currently operating or expected to be functional in the upcoming years. These experiments are expected to detect at least $\sim 10^4$ neutrino events from a galactic supernova explosion in the coming years~\cite{Mirizzi:2015eza,Horiuchi:2018ofe}, boosting our ability to probe neutrino physics and supernova mechanisms with unprecedented precision.

Inside the core of a core-collapse supernova, an enormous number of neutrinos~($\sim 10^{58}$) are generated through several processes in tens of seconds, carrying away 99\% of the total emitted energy~\cite{Janka:2017vlw,Mirizzi:2015eza}. These neutrinos experience highly dense and rapidly varying matter densities while escaping the supernova, which drastically influence the neutrino flavor evolution. 
Due to standard neutrino–matter interactions, active neutrinos, particularly electron neutrinos, undergo multiple adiabatic flavor transitions as they propagate from the dense core to the stellar envelope~\cite{Dighe:1999bi}. These transitions are sensitive to the neutrino mass ordering~\cite{Dasgupta:2008my} that affects the outgoing electron neutrino flux. As a result, the neutrino signal from a future supernova explosion may play a crucial role in establishing the neutrino mass ordering. However, physics beyond the Standard Model (BSM) can potentially modify the picture~\cite{Nunokawa:1997ct,Esteban-Pretel:2007zkv,Dighe:2017sur,deGouvea:2019goq,Chen:2022kal,Tang:2020pkp,Jana:2022tsa,Jana:2023ufy,dosSantos:2023skk,Carenza:2023old,Sen:2024fxa,Jana:2024lfm}. In this work, we investigate the impact of scalar-mediated non-standard interactions (SNSI) between neutrinos and background matter on supernova neutrino flavor evolution. These interactions modify the general Dirac mass term for neutrinos and can alter their propagation through matter~\cite{Ge:2018uhz}. SNSI effects have previously been explored in the context of solar and accelerator neutrino experiments~\cite{Babu:2019iml,Medhi:2021wxj,Singha:2023set,ESSnuSB:2023lbg,Sarker:2023qzp,Gupta:2023wct,Denton:2024upc,Dutta:2024hqq}. This work investigate the impact of flavor conserving SNSI on the supernova neutrino events in the upcoming neutrino experiments Deep Underground Neutrino Experiment (DUNE) and Hyper-Kamiokande (Hyper-K).

DUNE~\cite{DUNE:2020jqi,DUNE:2020ypp} and Hyper-K~\cite{Hyper-Kamiokande:2018ofw} are among the leading next-generation neutrino experiments. DUNE consists of a Liquid Argon Time Projection Chamber~(LArTPC) of 40 kt
located at the Sanford Underground Research Facility~(SURF) in the USA. It will receive a muon neutrino beam from the accelerator at Fermilab, situated at a distance of approximately 1300 km from SURF. The primary goal of this experiment is to resolve the unknown issues in neutrino oscillation such as leptonic CP violation and neutrino mass ordering~\cite{DUNE:2020jqi}. In addition to its accelerator-based program, DUNE is capable of detecting neutrinos from core-collapse supernovae, thanks to its low energy threshold~($\sim 5$~MeV)~\cite{DUNE:2020ypp}. The large volume of the detector and excellent efficiency make it a promising observatory for capturing supernova neutrino signals. Hyper-K is the successor of the Super-Kamiokande detector based in the Kamioka mine in Japan. It will eventually consist of two identical water Cherenkov detectors of 187 kt each, which includes a proposed second detector complex based in Korea~\cite{Hyper-Kamiokande:2018ofw}. The energy threshold to detect electron antineutrino events is as small as several MeV, making it ideal for supernova neutrino observations~\cite{Abe:2011ts}. We estimate the expected supernova neutrino event rate these two detectors would observe from future core-collapse supernovae and how the possible scalar-mediated interaction influences the event distributions.

The paper is organized as follows: In section~\ref{sec:SNSI-formalism}, we present the theoretical framework of scalar-mediated non-standard interactions and describe how these interactions modify the neutrino propagation Hamiltonian. Section~\ref{sec:flav_evolution} demonstrates the flavor evolution of the neutrino propagating from the supernova core to the surface and how the presence of scalar-mediated interaction modifies the propagation. In Section~\ref{sec:Events}, we compute the expected event distributions of supernova neutrinos at DUNE and Hyper-K and discuss our results. Finally, we conclude in section~\ref{sec:Summary}. In appendix \ref{app:A1}, we show the adiabaticity condition for neutrino propagation in the presence of scalar NSI.

%===================================================
\section{Theoretical formalism of SNSI}
\label{sec:SNSI-formalism}
%==================================================

Neutrinos interact weakly with matter particles as they propagate through a medium like the Earth, the Sun or an astrophysical object such as a supernova. In the standard model (SM), neutrino-matter interactions are mediated by the $W$ and $Z$ vector bosons, corresponding to the charged-current~(CC) and neutral-current~(NC) interactions, respectively. Neutrino propagation for neutrino energy $E$ is governed by~\cite{PhysRevD.17.2369,Blennow:2013rca}
\begin{equation}
    {\cal H} = E\cdot  I+ H^{\rm flavor}_{\rm vac} \pm V_{\rm SI}
\end{equation}
where $V_{\rm SI}= {\rm Diag}[V_{\rm CC},0,0]$ and in terms of the Fermi coupling and the number density of electrons, $V_{\rm CC} = \sqrt{2}G_F N_e$.
$I$ in the first term denotes an identity matrix. 
The $+(-)$ is for (anti-)neutrinos.
The SM vacuum flavor Hamiltonian is
\begin{equation}
    H^{\rm flavor}_{\rm vac} \simeq  U\cdot {\rm Diag} \Biggl[\frac{m_1^2}{2E},\frac{m_2^2}{2E},\frac{m_3^2}{2E}  \Biggr] \cdot U^\dagger=\frac{M M^\dagger}{2E}\,,
\end{equation}
where neutrino flavor and mass eigenstates are related by, e.g.,  
$|\nu_e\rangle = \sum_{i=1,2,3}U_{ei}|\nu_{i}\rangle$, for mass eigenstates
$|\nu_{i}\rangle$. For SM (active) neutrinos, this means that 
\begin{equation}
 M_{\alpha\beta} = \left[U\cdot\mathrm{Diag}(m_1,m_2,m_3)\cdot U^\dagger\right]_{\alpha\beta}\,,  
 \label{eq:vac_Ham}
\end{equation}
for the neutrino mixing matrix $U$.
%, and $m_i$ denotes the neutrino mass states.

Various neutrino mass models propose new beyond the standard model~(BSM) interactions between neutrinos and charge fermions, mediated by a new vector boson or a scalar particle. The vector-mediated interactions are known as non-standard interactions (NSI)~\cite{Farzan:2017xzy,Proceedings:2019qno,Esteban:2018ppq,Agarwalla:2021zfr}, which generate interactions in addition to the standard CC and NC interactions. In this work, we consider new interactions mediated by a scalar particle $\phi$. 

Theoretically, neutrino mass models in which the SM particle content is extended with a new scalar singlet $\phi$ and right-handed neutrinos can introduce such interactions ~\cite{Babu:2019iml}. The new scalar particle couples to both neutrinos and charged fermions.
The effective interaction term for SNSI can be written as
\begin{align}
\mathcal{L}^{{\rm SNSI}}_{{\rm eff}} = \frac{y_f Y_{\alpha\beta}}{m^2_\phi}[\bar{\nu}_\alpha(p_3)\nu_\beta(p_2)][\bar{f}(p_1)f(p_4)]\,,
\end{align}
where $y$ and $Y_{\alpha\beta}$ show the coupling strength of charged fermions $f$~($f = e,u,d$) and  SM neutrinos $(\nu_\alpha = \nu_e,\nu_\mu, \nu_\tau)$ with the scalar mediator $\phi$, respectively. The mediator mass $m_\phi$ and the couplings $Y_{\alpha\beta}$, $y_f$ are constrained by various observables like the leptonic decay of mesons~\cite{Pasquini:2015fjv}, big
bang nucleosynthesis (BBN)~\cite{Huang:2017egl,Babu:2019iml}, coherent scattering~\cite{Farzan:2018gtr}, neutrino events from 1987A~\cite{Heurtier:2016otg,Fiorillo:2022cdq}, fifth force experiments~\cite{Babu:2019iml}, and many others. The expected constraints on $Y_{\alpha\beta}$ and $m_\phi$ from future supernova events have been discussed in refs.~\cite{Akita:2022etk,Lazar:2024ovc,Telalovic:2024cot,Akita:2023iwq}.

The new scalar-mediated interaction modifies the neutrino mass term in the Dirac equation so that
\begin{equation}
  M_{\alpha\beta } \to M_{\alpha\beta }^{\rm eff} =  M_{\alpha\beta }
+ \frac {\sum_f n_f y_f Y_{\alpha\beta}}{m^2_\phi}  
\end{equation}
when non-relativistic fermions with number densities $n_f$ dominate the effective mass correction, which we assume here. Note that, inside the supernova core, electrons in the background medium can be relativistic, for which the correction term is proportional to $n_e^{2/3}$~\cite{Babu:2019iml}. However, since the scalar-mediated interactions are primarily governed by the nucleons inside the core, considering relativistic electrons does not change our results qualitatively. Based on ref. \cite{Babu:2019iml}, we estimate that the scalar NSI contribution from relativistic electrons in the supernova core is suppressed by almost three orders of magnitude compared to that from non-relativistic nucleons.
The effective neutrino propagation Hamiltonian in the presence of SNSI is written as
\begin{equation}
\mathcal H
\approx E\cdot I +
\frac {M^{{\rm eff}}{M^{{\rm eff}}}^\dagger}{2 E}
\pm V_{\rm SI} \,.
\label{eq:Hs}
\end{equation}

For a turbulent background medium, like in a supernova, the strength of the new interaction depends on the local fermion density. To understand the effective impact of the varying matter density in a supernova, following ref.~\cite{Ge:2018uhz}, we redefine the modified mass matrix as 
\begin{equation}
M^{\rm eff} 
=
M
+ \delta M
%\equiv
%M_{\rm re}
%+ \delta  M(n_{f,s}) \frac {n_f - n_{f,s}}{n{f,s}} \,,
\equiv
 M_{\rm re}
+ \delta  \tilde{M} \frac {\rho - \rho_s}{\rho_s} \,,
\label{eq:scaling_density}
\end{equation}
where $M_{\rm re} = M+\delta \tilde{M}$, $\rho$ is the local matter density converted from the fermion number density, and $\rho_s$ is a fixed scaling density, generally taken to be the average density of the medium. We use $\rho_s=10^5$ g/cc.
$\delta\tilde{M}$ denotes the matrix $\delta M$ computed at the scaling density $\rho_s$, such that $\delta M = \delta \tilde{M}\frac{\rho}{\rho_s}$. The matrix, $\delta \tilde{M}$, can be parameterized in terms of dimensionless quantities $\eta_{\alpha\beta}$ after scaling it with $\sqrt{|\Delta m^2_{31}|}\equiv\sqrt{|m^2_3-m^2_1|}$,
\begin{equation}
\delta  \tilde{M}
\equiv
\sqrt{|\Delta m^2_{31}|}
\left\lgroup
\begin{matrix}
\eta_{ee}     & \eta_{e \mu}    & \eta_{e \tau}   \\
\eta_{\mu e}  & \eta_{\mu \mu}  & \eta_{\mu \tau} \\
\eta_{\tau e} & \eta_{\tau \mu} & \eta_{\tau \tau}
\end{matrix}
\right\rgroup \,,
\label{eq:dM}
\end{equation}
where $\eta_{\alpha\beta} \propto \frac{\sum_f n_{f,s} y_f Y_{\alpha\beta}}{m^2_\phi\sqrt{|\Delta m^2_{31}|}}$, with $n_{f,s}$ denoting the average number density of the fermion $f$. 
This work focuses on scalar-mediated non-standard interactions of supernova neutrinos. We consider the SNSI correction term $\delta M$ to be less than 5 MeV~\cite{Smirnov:2019cae,Babu:2019iml}. We also assume neutrino coupling $y_\nu$ to be small $y_{\alpha\beta}\lesssim 10^{-10}$, as a large coupling may lead to a significant contribution from neutrino self-interactions~\cite{Fiorillo:2023cas,Fiorillo:2023ytr}.

Unlike the vector-mediated interaction, the SNSI does not flip sign for antineutrinos relative to neutrinos, as the interaction terms are invariant under charge-conjugation. Consequently, the impact of SNSI on the matrix $\delta M$ remains the same for antineutrinos and neutrinos. As usual, in the neutrino and antineutrino propagation Hamiltonian in eq.~(\ref{eq:Hs}), the sign of the interaction matrix $V_{\rm SI}$ flips.

%===================================================
\section{Neutrino flavor evolution in core-collapse supernova in the presence of flavor conserving SNSI}
\label{sec:flav_evolution}
%=================================================

During the various stages of a core-collapse supernova, all three neutrino flavors ($\nu_e$, $\nu_\mu$, and $\nu_\tau$) are generated through different physical processes (see, e.g., ref.~\cite{Janka:2017vlw}).
During the neutronization-burst phase, which occurs inside the core of the star during tens of milliseconds after the core bounce, electrons in the medium are captured by the free protons $(e^-+p\rightarrow n + \nu_e)$ to produce a large number of electron neutrinos. As a result, the $\nu_e$ flux dominates over the other two SM neutrino flavors in this stage. The generated $\nu_e$ flux remains trapped in highly dense post-shock matter before they freely stream outwards once the surrounding matter becomes sufficiently transparent as the density drops~\cite{Janka:2017vlw}. Shortly after the neutronization burst, the loss of electron–lepton number due to the escape of the $\nu_e$ flux leads to the thermal production of positrons. This accelerates the production of $\bar{\nu}_e$ and the two other neutrino and antineutrino flavors~($\nu_x, \bar{\nu}_x = \nu_\mu, \nu_\tau,\bar{\nu}_\mu,  \bar{\nu}_\tau$)  via pair-annihilation ($e^++e^-\rightarrow \nu+\bar{\nu}$) and nucleon-nucleon bremsstrahlung ($N+N'\rightarrow N+N'+\nu+\bar{\nu}$). 
Additionally, positron capture on neutrons ($e^++n\rightarrow p + \bar{\nu}_e$) contributes to the production of $\bar{\nu}_e$ at the end of the neutronization phase and in the post-neutronization phase. 
In the post-neutronization phase, since there is a high density of neutrinos and comparatively low matter density due to the expansion of the shock wave, collective neutrino oscillations from neutrino self-interactions play an important role in the neutrino flavor evolution~\cite{Duan:2010bg,Chakraborty:2016yeg}.
However, in this work, we place a particular emphasis on
the impact of neutrino-matter interactions in the flavor evolution, most important in the neutronization-burst phase when the effects from collective neutrino oscillations are relatively suppressed as compared to matter-induced flavor evolution in the high matter densities inside the core.  A full treatment that also
considers the impact  of collective  oscillations in neutrino propagation is beyond scope of this work. We note that
the $\bar{\nu}_e$ flux in the neutronization phase is very small compared to the $\nu_e$ flux. The $\bar{\nu}_e$ flux may be large enough for detection by large volume experiments such as Hyper-K~\cite{deGouvea:2019goq}, but only if supernova neutrino luminosity is large enough, and the supernova is relatively close, as discussed below.

The time- and energy-dependent distribution of the unoscillated neutrino flux emitted during a supernova explosion can be expressed as
\begin{equation}
\Phi_\nu(E,t) = \frac{L_\nu(t)}{\langle E_\nu \rangle}  \phi(E)\,,
\label{eq:flux}
\end{equation}
where $L_\nu(t)$ is the neutrino luminosity at time $t$ after the core bounce, and $\langle E_\nu \rangle$ is the corresponding average neutrino energy. The quantity $\phi(E)$ is the neutrino spectrum, which often represented in terms of a modified Fermi-Dirac distribution~\cite{Keil:2002in}
\begin{equation}
\phi(E)=\frac{1}{\langle E_\nu \rangle}\frac{(\alpha+1)^{(\alpha+1)}}{\Gamma (\alpha+1)}\left(\frac{E}{\langle E_\nu \rangle}\right)^{\alpha}{\rm exp}\left[-(\alpha+1) \frac{E}{\langle E_\nu \rangle}\right]\,,
\label{spectra_ch1}
\end{equation}
where $\alpha$ is the spectral pinching parameter. In this work, we consider neutrino luminosities and the average energy from the simulation result from ref.~\cite{Garching} for a $25 M_\odot$ progenitor model. 

The neutrino flux $\Phi_\nu (E,t)$ generated within the supernova core for each flavor is significantly modified by the neutrino flavor mixing effect. In the dense inner regions of the exploding star, where matter densities can reach up to $\sim 10^{14}$ g/cm$^3$, neutrino-matter interactions—particularly the CC interactions involving electron neutrinos—play a dominant role in the neutrino mixing, thus affecting the neutrino flavor evolution. In addition, possible scalar-mediated non-standard interactions can further influence the neutrino flavor evolution. As neutrinos propagate through the varying matter density of the stellar environment, the effective mixing between the neutrino flavor states varies with local matter density and neutrino energy. 
To have a better understanding of the effective neutrino mixing inside the star during a supernova explosion, we diagonalize the total neutrino propagation Hamiltonian, as defined in eq.~(\ref{eq:Hs}), $\mathcal{H}_{\rm diag} = {U^{\rm eff}}^\dagger \cdot \mathcal{H}\cdot U^{\rm eff}$. The effective mixing matrix $U^{\rm eff}$ determines the mixing between the neutrino flavor states and the effective mass states, $\nu_{i,m}$~($i=1,2,3$). 

We consider a radial matter density profile that decreases from $\sim10^{14}$ g/cm$^3$ at the stellar core to zero at the envelope~\cite{Tang:2020pkp, Jana:2024lfm}. 
The matter potential is computed using the electron number density profile described in Refs.~\cite{Tang:2020pkp, Jana:2024lfm}. In the standard model case, where outgoing neutrinos interact with the highly dense matter~($\rho>10^5$ g/cm$^3$) inside the core via CC and NC interactions, the effective mixing matrix elements for electron neutrinos take the values $|U^{\rm eff}_{e3}|\simeq  1$, $|U^{\rm eff}_{e2}|\simeq |U^{\rm eff}_{e1}|\simeq 0$ for normal mass ordering~(NMO) of neutrinos ($m_1 < m_2 < m_3)$~\cite{Dighe:1999bi}.  In other words, electron neutrino flux produced inside the supernova primarily aligns with the $\nu_{3,m}$ state. 
It is verified that the propagation of the neutrino mass states through the varying matter density in the supernova maintains adiabaticity and is incoherent, such that $(\nu_{i,m})$ state reaches the surface of the supernova as the $\nu_i$ state. This means that the flux of $\nu_{e}$, $\Phi(\nu_e)\simeq \Phi (\nu_{3,m})$ reaches the surface as $\Phi(\nu_3)$. The resulting expression for the $\Phi(\nu_e)$ flux at the supernova surface, assuming NMO in the standard model (SI), is given by
\begin{equation}
    \Phi(\nu_e) = |U_{e3}|^2 \Phi_c(\nu_e) + (1-|U_{e3}|^2)\, \Phi_c(\nu_x)\,\,\,\,\,\,...\,\,{\rm for\ NMO(SI)},
    \label{eq:flux_nu_e}
\end{equation}
where $\Phi_c(\nu_{\alpha})$ denotes the $\nu_\alpha$ flux at the supernova core estimated using eq.~(\ref{eq:flux}). The last term in eq.~(\ref{eq:flux_nu_e}) accounts for the contribution from the other two flavors of the neutrino, assuming $\Phi_c(\nu_\mu) = \Phi_c(\nu_\tau) = \Phi_c(\nu_x)$. For the inverted mass ordering~(IMO) of neutrino ($m_3< m_1<m_2$), $\nu_e$ aligns with the $\nu_{2,m}$ state, and the fraction of electron neutrinos at the supernova surface is
\begin{equation}
    \Phi(\nu_e) = |U_{e2}|^2 \Phi_c(\nu_e) + (1-|U_{e2}|^2) \Phi_c(\nu_x)\,\,\,\,\,\,...\,\,{\rm for\ IMO(SI)}\,.
    \label{eq:flux_nux}
\end{equation}

\begin{figure}[tbp]
    \centering
    \includegraphics[width=0.99\textwidth]{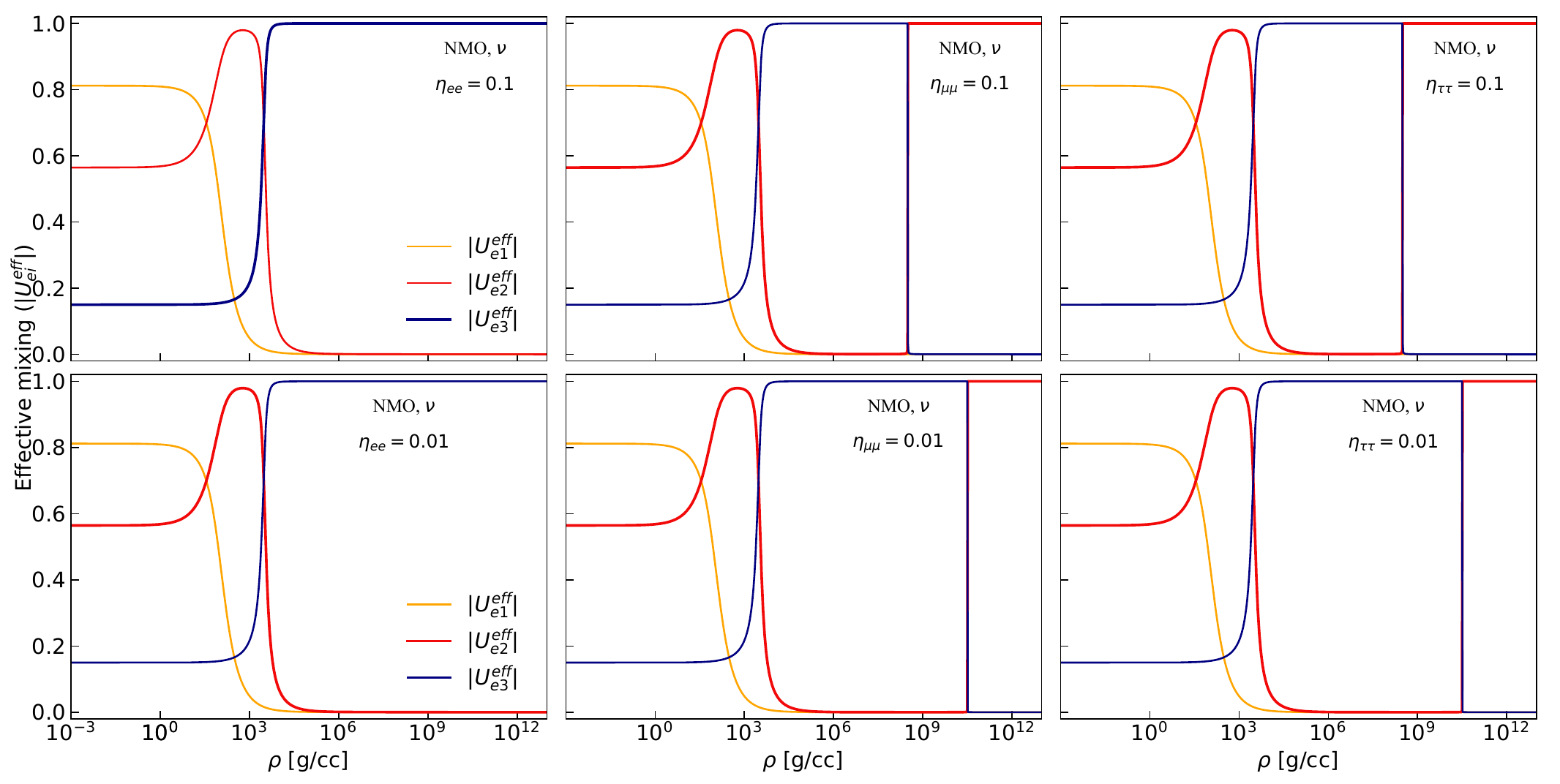}
    \caption{For normal mass ordering, the effective mixing elements $|U_{e1}^{\rm eff}|$, $|U_{e2}^{\rm eff}|$, and $|U_{e3}^{\rm eff}|$ as a functions of matter density in the presence of scalar mediated NSI parameters, $\eta_{ee}$ (left column), $\eta_{\mu\mu}$ (middle column), and $\eta_{\tau\tau}$ (right column). The top and bottom rows correspond to the strength of the SNSI parameters $\eta_{\alpha\alpha} = 0.1$ and $\eta_{\alpha\alpha} = 0.01$, respectively. We consider neutrino energy $E = 10$~MeV in the plot. The value of oscillation parameters used to compute the mixing matrix elements are taken from NuFit-6~\cite{Esteban:2024eli}.}
    \label{fig:mass-mix-nu-nmo}
\end{figure}

\begin{figure}[tbp]
    \centering
    \includegraphics[width=0.99\textwidth]{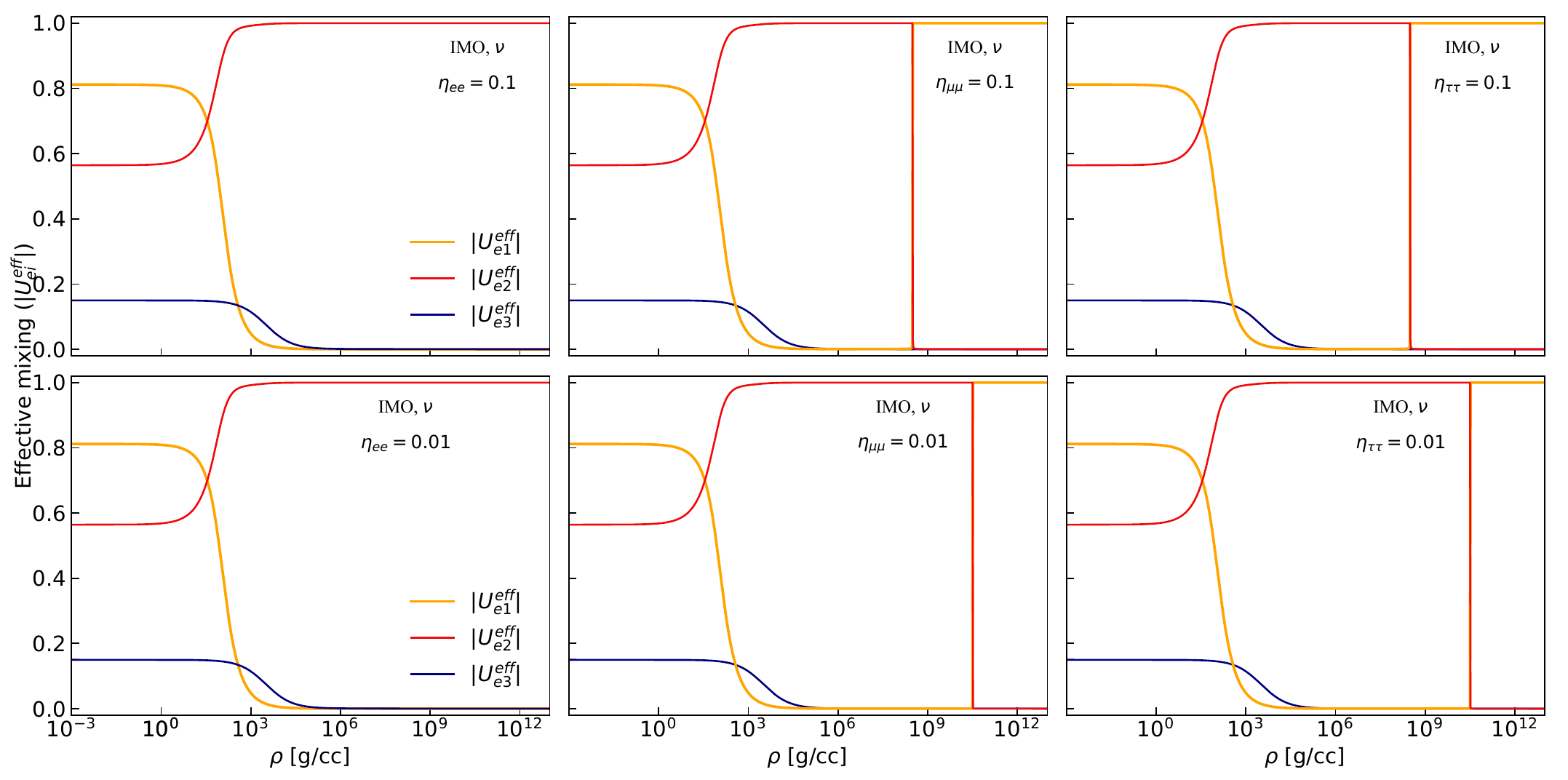}
    \caption{Same as fig.~\ref{fig:mass-mix-nu-nmo}, but with inverted mass ordering of neutrino masses.}
    \label{fig:mass-mix-nu-imo}
\end{figure}

We now investigate the impact of scalar-mediated non-standard interactions on neutrino flavor evolution in the supernova environment. Specifically, we focus on flavor-diagonal SNSI parameters: $\eta_{ee}$, $\eta_{\mu\mu}$, and $\eta_{\tau\tau}$.
For our analysis, we consider two benchmark values $\eta_{\alpha\alpha} = 0.1$ and $0.01 ~(\alpha = e,\mu,\tau)$ introducing one parameter at a time. We checked that these values lies within the bounds allowed by existing constraints on SNSI interaction~\cite{Babu:2019iml}. We also checked that neutrino propagation is adiabatic inside the supernova even in the presence of SNSI (see appendix~\ref{app:A1}). 
 In fig.~\ref{fig:mass-mix-nu-nmo}, we show the evolution of the effective mixing matrix elements involving $\nu_e$ ($U^{\rm eff}_{e1}$, $U^{\rm eff}_{e2}$, $U^{\rm eff}_{e3}$) as a function of matter density relevant for the supernova environment for neutrino energy $E=10$ MeV for NMO. The three columns show the presence of the SNSI parameters $\eta_{ee}$, $\eta_{\mu\mu}$ and $\eta_{\tau\tau}$ one at a time. The top row shows results with SNSI strength $\eta_{\alpha\alpha} = 0.1$, while the bottom row corresponds to the results with $\eta_{\alpha\alpha} = 0.01$.
  In the presence of the SNSI parameter $\eta_{ee}$, we find that the effective mixing remains the same as in the standard case inside the supernova core where $\rho\gtrsim 10^{11}$ g/cc, {\it i.e.}, $|U^{\rm eff}_{e3}|^2\simeq 1$, $|U^{\rm eff}_{e2}|^2\simeq|U^{\rm eff}_{e1}|^2 \simeq 0$. However, in the presence of SNSI parameters $\eta_{\mu\mu}$ or $\eta_{\tau\tau}$, we observe sharp changes in the effective mixing parameters compared to the standard case. For $\eta_{\mu\mu}= 0.1$ or $\eta_{\tau\tau}=0.1$, $|U^{\rm eff}_{e2}|\simeq 1$ and $|U^{\rm eff}_{e1}|=|U^{\rm eff}_{e3}|\simeq 0$ at the supernova core ($\rho \gtrsim 10^{11}$ g/cc). 
This happens because of the sharp transition between the effective mixings $|U^{\rm eff}_{e3}|$ and $|U^{\rm eff}_{e2}|$
 as seen in fig.~\ref{fig:mass-mix-nu-nmo}~\footnote{In fig.~\ref{fig:mass-mix-nu-nmo}, $|U^{\rm eff}_{e2}|$ increases from 0 to 1 (while $|U^{\rm eff}_{e3}|$ decreases from 1 to 0) at a density $\rho =\rho_{\rm res}\simeq 4\times 10^8$ g/cc~($4\times 10^{10}$ g/cc) in the presence of $\eta_{\mu\mu}$ or $\eta_{\tau\tau}$ with strength 0.1~(0.01). This leads to a transition between $\nu_{3,m}$ and $\nu_{2,m}$ and results in $\nu_e\simeq\nu_{3,m}$ to $\nu_e\simeq \nu_{2,m}$ at $\rho> \rho_{\rm res}$.}, where the transition occurs for $\rho<10^{11} $ g/cc. As long as the sharp transition occurs outside of the core, essentially all of the $\nu_e$ flux produced in the neutronization phase is in the $\nu_{2,m}$ state. Now, at the supernova core, $\nu_e$ aligns with the $\nu_{2,m}$~($\nu_e\simeq \nu_{2,m}$).
In fig.~\ref{fig:mass-mix-nu-imo}, we show the same as in  fig.~\ref{fig:mass-mix-nu-nmo}, but for the inverted mass ordering of neutrinos. Similar to the NMO case, we observe a  flip between the $|U^{\rm eff}_{e1}|$ and $|U^{\rm eff}_{e2}|$ in presence of SNSI parameters $\eta_{\mu\mu}$ or $\eta_{\tau\tau}$. Now, $\nu_e$ aligns with the $\nu_{1,m}$~($\nu_e\simeq\nu_{1,m}$) state inside the supernova core.
  
In figs.~\ref{fig:mass-mix-nu-nmo} and \ref{fig:mass-mix-nu-imo}, we took the evolution of the effective mixing given the matter density inside the supernova, considering the strength of SNSI parameters as low as 0.01. As we go toward lower values of the SNSI parameters, the transition $\nu_{e}\simeq \nu_{2,m}$~($\nu_{e}\simeq \nu_{1,m}$) from $\nu_{e}\simeq \nu_{3,m}$~($\nu_{e}\simeq \nu_{2,m}$) states for NMO~(IMO), which we call the ``SNSI resonance", occurs at a higher density. When the SNSI parameter $\eta_{\mu\mu}$ or $\eta_{\tau\tau}$ is too small, the resonance density is above the critical density $\rho_c = 10^{14}$ g/cc, so resonance due to SNSI does not occur within the supernova.  As noted above, we also required that the SNSI resonance occur outside of the core so that all $\nu_e$ in the neutronization phase align with a singe mass state $\nu_{i,m}$.

To estimate the minimum allowed value of the SNSI parameter for which the new resonance occurs within the supernova matter profile,
we derive an expression for the resonance condition %for the matter density at which the new resonance occurs 
using a two-flavor approximation. Considering the $(e,\mu)$ pair in the Hamiltonian, and neglecting the elements from the vacuum Hamiltonian\footnote{We checked that the value of elements of the vacuum Hamiltonian are several orders smaller than the interaction part of the Hamiltonian at the relevant matter densities.}
\begin{align}
    \mathcal{H}_{2\times2} = \frac{1}{2E}
    \left\lgroup
    \begin{matrix}
    2EV_{\rm CC} & M_{e\mu}\delta M_{\mu\mu}\\
    M^\ast_{e\mu}\delta M_{\mu\mu} &  (\delta M_{\mu\mu})^2
    \end{matrix}
    \right\rgroup\,,
\end{align}
In the above equation, $M_{\alpha\beta}~(\delta M_{\alpha\alpha})$ is the element of the vacuum Hamiltonian as defined in eq.~(\ref{eq:vac_Ham}) (eq.~(\ref{eq:scaling_density})). Diagonalizing the Hamiltonian, we derive the resonance condition
\begin{align}
    2EV_{\rm CC} = %%%%%2 M_{\mu\mu} 
    (\delta M_{\mu\mu})^2\,.
    \label{eq:res_cond}
\end{align}
The above resonance condition for neutrinos gives the expression for the density at resonance,
\begin{align}
    \rho_{\rm res} = \frac{2\sqrt{2}G_F Y_e E}{m_n |\Delta m^2_{31}|\eta^2_{\mu\mu}}\rho^2_s\,,
    \label{eq:rho_res}
\end{align}
where $m_n$ is nucleon mass, and $Y_e$ is the electron fraction. As mentioned earlier, we take $\rho_s=10^{5}$ g/cc in this work. The resonance density depends on $\eta_{\mu\mu}^2$, so it is independent of the sign of $\eta_{\mu\mu}$. Since the new resonance density also depends on the absolute value of the mass-squared difference, we observe the new resonance for both the mass orderings of neutrinos. For NMO, $m_2$ and $m_3$ are the closest mass states, so we see a transition between $|U^{\rm eff}_{e2}|$ and $|U^{\rm eff}_{e3}|$.  However, for IMO, $m_2$ and $m_{1}$ are the closest mass states, and we see a transition between $|U^{\rm eff}_{e1}|$ and $|U^{\rm eff}_{e2}|$. For antineutrinos, $V_{\rm CC}$ changes sign and the resonance condition in eq.~(\ref{eq:res_cond}) is not valid anymore, even for negative SNSI parameters, as it appears in the quadratic form in eq.~(\ref{eq:rho_res}). As a result, we do not observe any new resonance due to the SNSI in the case of antineutrinos.

\begin{figure}
    \centering
    \includegraphics[width=0.49\linewidth]{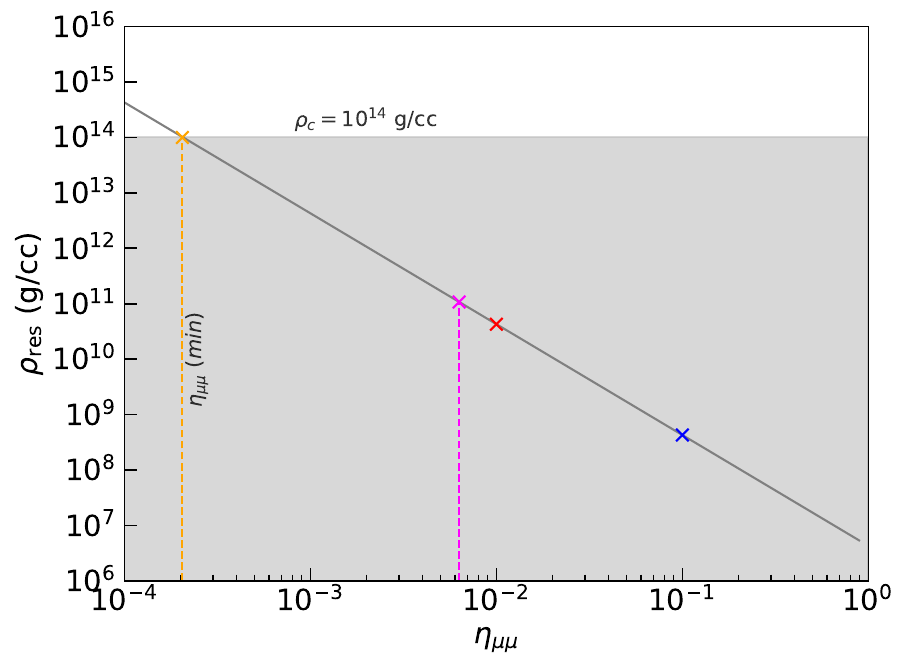}
    \includegraphics[width = 0.49\linewidth]{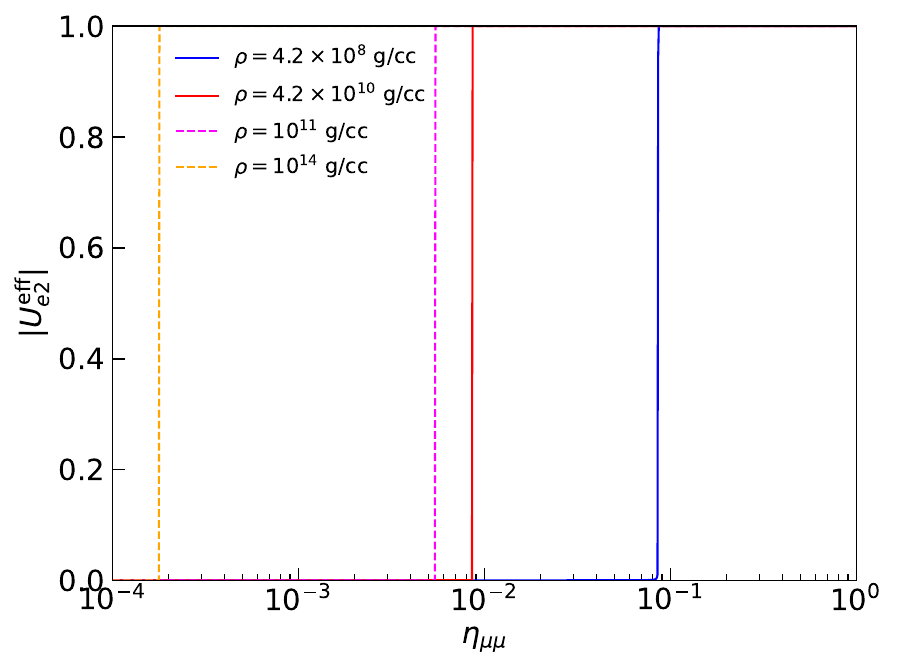}
    \caption{\textbf{Left:}~For neutrinos, the resonance density $\rho_{\rm res}$~(eq.~(\ref{eq:rho_res})) as a function of the SNSI strength~$\eta_{\mu\mu}$ estimated analytically with two flavor approximation. The shaded region shows the possible matter densities inside the supernova~($\rho<10^{14}$ g/cc). The orange vertical line corresponds to the minimum value of $\eta_{\mu\mu}$, which keeps the new resonance inside the supernova. The magenta vertical line depicts the $\eta_{\mu\mu}$ for which resonance occurs at $\rho \simeq 10^{11}$ g/cc, the density at the core boundary. The red and blue dots show the resonance densities that correspond to the $\eta_{\mu\mu} = 0.01$ and $\eta_{\mu\mu} = 0.1$, respectively. \textbf{Right:} Effective mixing $|U^{\rm eff}_{e2}|$ as function of the $\eta_{\mu\mu}$ for four benchmark value of the matter density inside the supernova: $\rho = 4.2\times 10^8$ g/cc ($\rho_{res}$ for $\eta_{\mu\mu} = 0.1$), $\rho = 4.2\times 10^{10}$ g/cc ($\rho_{res}$ for $\eta_{\mu\mu} = 0.01$), $\rho = 4.2\times 10^{11}$ g/cc ($\rho_{res}$ for $\eta_{\mu\mu} = 0.006$), and $\rho = 10^{14}$ g/cc ($\rho_{res}$ for $\eta_{\mu\mu} = 2\times 10^{-4}$ g/cc). $|U^{\rm eff}_{e2}|$ is computed numerically with full three-flavor formalism. We consider neutrino energy $E= 10$ MeV in both the panels.}
    \label{fig:Resonance}
\end{figure}

To check the validity of the two-flavor approximation to derive eq. (\ref{eq:rho_res}), we plot $\rho_{\rm res}$ as a function of $\eta_{\mu\mu}$ in the left panel of fig.~\ref{fig:Resonance} using $E= 10$ MeV. The blue and red dots correspond to $\eta_{\mu\mu} =0.1$ and $0.01$, for which resonance densities are $\rho_{\rm res}\simeq 4.2\times 10^{8}$ g/cc and $4.2\times 10^{10}$ g/cc. 
In the right panel, we show the evolution of $|U^{\rm eff}_{e2}|$ at these two densities and at same neutrino energy, as indicated by the blue and red lines. We observe that the transition occurs at $\eta_{\mu\mu}\simeq 0.1$ and $\eta_{\mu\mu} \simeq 0.01$, respectively, which correspond to the blue and red dots in the left panel. There is a small deviation from the $\eta_{\mu\mu}$ equal to $0.1$ and $0.01$ due to the two-flavor approximation. 

To have all the electron neutrinos produced during the neutronization burst phase in the new effective mass states~($\nu_{2,m}$~($\nu_{1,m}$) state for NMO~(IMO)), the SNSI resonance needs to occur outside the core of the supernova inside of which the neutronization phase occurs. Considering $\rho \simeq 10^{11}$ g/cc at the boundary of supernova core~\cite{Jana:2024lfm}, the dashed magenta line shows the minimum value of $\eta_{\mu\mu}$~($\approx 0.006$) for which SNSI resonance for a 10 MeV neutrino occurs outside the supernova core. 
This means that for $\eta_{\mu\mu}\gtrsim 0.006$, the entire $\nu_e$ flux produced during the neutronization phase aligns with the $\nu_{2,m}$~($\nu_{1,m}$) for NMO~(IMO).
We checked that for $E= 40$ MeV, the maximum neutrino energy we consider in this work, this limit becomes $\eta_{\mu\mu}\gtrsim 0.01$. However, neutrino flux at such high energies is so small  that the the number of events from these energies is negligible.
The gray shaded region in the left panel shows the possible densities inside the supernova~($\rho~<~10^{14}$ g/cc).
The minimum value of $\eta_{\mu\mu}$ that produces the new resonance  inside the supernova is $\eta_{\mu\mu}({\rm min}) \simeq 2\times 10^{-4}$. We checked that our two-flavor approximation of $\eta_{\mu\mu}({\rm min})$ matches the exact three-flavor scenario, as shown by the orange line in the right panel.

The resulting $\nu_e$ flux at the surface, $\Phi_{\nu_e}$ at the supernova envelope is written as
\begin{align}
 \Phi(\nu_e) &= |U_{e2}|^2 \Phi_c(\nu_e) + (1-|U_{e2}|^2)\, \Phi_c(\nu_x)\,\,\,\,\,\,...\,\,{\rm for\ NMO}\\
 \Phi(\nu_e) &= |U_{e1}|^2 \Phi_c(\nu_e) + (1-|U_{e1}|^2)\, \Phi_c(\nu_x)\,\,\,\,\,\,...\,\,{\rm for\ IMO}\,.
\end{align}
For the antineutrino case, evolution of the neutrinos flavor does not change in the presence of flavor-conserving SNSI parameters, neither in the NMO nor the IMO scenario. The $\bar{\nu}_e$ aligns with the $\bar{\nu}_{1,m}$ states at the supernova core for NMO. For the IMO case, $\bar{\nu}_e$ aligns with $\bar{\nu}_{3,m}$ state. The fraction of $\bar{\nu}_e$ at the surface of the supernova are
\begin{align}
    \Phi(\bar{\nu}_e) &= |U_{e1}|^2\,\Phi_c (\bar{\nu}_{e})+(1-|U_{e1}|^2)\Phi_c(\bar{\nu}_{x})\,\,\,\,\,\,\,\, {\rm ...\,\,\,for\ NMO}\\
    \Phi (\bar{\nu}_e) &= |U_{e3}|^2\Phi_c(\bar{\nu}_{e})+(1-|U_{e3}|^2)\Phi_c(\bar{\nu}_{x})\,\,\,\,\,\,\,\,\, {\rm ...\,\,\,for\ IMO}\,.
\end{align}

\begin{table}[tbp]
    \centering
    \begin{tabular}{ |c|c|c| }
		\hline\hline
		& NMO & IMO \\ 
		\hline
		$\nu_e$	(SI case) & $\nu_e\approx\nu_{3,m}$ & $\nu_e\approx\nu_{2,m}$ \\  
		\hline
	$\nu_e$ ($\eta_{\mu\mu}$ or $\eta_{\tau\tau}$) & $\nu_e\approx\nu_{2,m}$ & $\nu_e\approx\nu_{1,m}$ \\
		\hline\hline
		$\bar{\nu}_e$ (SI case) & $\bar{\nu}_e\approx\bar{\nu}_{1,m}$  & $\bar{\nu}_e\approx\bar{\nu}_{3,m}$\\
		\hline
		$\bar{\nu}_e$ ($\eta_{\mu\mu}$ or $\eta_{\tau\tau}$) & $\bar{\nu}_e\approx\bar{\nu}_{1,m}$  & $\bar{\nu}_e\approx\bar{\nu}_{3,m}$\\
		\hline\hline
	\end{tabular}
    \caption{The effective mass states to which $\nu_e$ and $\bar{\nu}_e$ align in supernova core considering the standard CC interaction (SI) and in the presence of SNSI parameters $\eta_{\mu\mu}$ or $\eta_{\tau\tau}$.}
    \label{tab:mass_states}
\end{table}

In table~\ref{tab:mass_states}, we summarize our observation, tabulating the effective mass states in which $\nu_e$ and $\bar{\nu}_e$ align in the standard model interaction case~(SI case) and in the presence of SNSI parameters $\eta_{\mu\mu}$ and $\eta_{\tau\tau}$.

%====================================
\section{Supernova neutrino events in next-generation neutrino experiments}
\label{sec:Events}
%====================================

Neutrinos produced in a supernova explosion can be detected at the Earth's surface. The flux of neutrinos of flavor $\nu_\alpha$ observed at Earth from a supernova located at a distance $R$ is given by 
\begin{equation}
    f_{\nu_\alpha,\oplus}(E) = \frac{\Phi(\nu_\alpha,E)}{4\pi R^2}\,,
    \label{eq:flux_at_earth}
\end{equation}
taking the geometrical factor for a source into account. In this work, we consider the neutrino flux from a supernova explosion at a distance of 10 kpc from Earth.

For a detector volume  $V$ placed at Earth, the expected number of supernova neutrino events per unit time in each reconstructed energy~($E^{rec}$) bin is given by
\begin{equation}
\mathrm{\frac{d^2N_{\nu_\alpha}}{dt\,dE^{rec}}} = N \int_{E^{tr}_{min}}^{E^{tr}_{max}} dE^{tr}\,\,f_{\nu_\alpha ,\oplus}(E^{tr})\times \sigma (E^{tr})\times \epsilon (E^{rec},E^{tr})\,.
\end{equation}
Here, $N$ in the above equation denotes the total number of targets within the detector volume, $E^{tr}$ is the true energy of the incoming neutrinos, $f_{\nu_\alpha ,\oplus}$ is the incoming neutrino flux at Earth calculated using eq.~(\ref{eq:flux_at_earth}) and $\epsilon(E^{rec},E^{tr})$ represents the energy resolution function. For this work, we consider a Gaussian energy resolution function,
\begin{equation}
\epsilon(E^{rec},E^{tr}) = \frac{1}{\sqrt{2\pi \sigma}} \,{\rm exp}\left[-\frac{(E^{tr}-E^{rec})^2}{2\sigma}\right]\,.
\label{eq:energy_res}
\end{equation}

In this study, we estimate the supernova neutrino event rates at two upcoming next-generation detectors: DUNE and Hyper-K. For DUNE, we consider the detector of 40 kt fiducial mass, which consists of 4 modules of 10 kt LArTPC detectors~\cite{DUNE:2020ypp}. This is equivalent to $5.47\times 10^{32}$ Argon nuclei as target particles. We assume an incoming electron neutrino flux spanning an energy range of 2–40 MeV. The incoming $\nu_e$ flux interacts with Argon nuclei via CC interaction, $\nu_e$ + $^{38}$Ar$\rightarrow$ $^{40}$K$^*$ + $e^-$. We simulate the flux-averaged cross-section of the $\nu_e$ interaction with Ar nuclei using the Monte Carlo event generator MARLEY~\cite{marleyPRC,marleyCPC}. To estimate the neutrino flux, we use the neutrino luminosity and average energy for the $25M_\odot$ model to estimate the distribution of the unoscillated neutrino neutrino flux as a function of energy and time after the core bounce. For the time resolution, we divide our simulated data into seven bins with a width of 5 ms for times up to 35 ms after the core bounce, where the $\nu_e$ flux dominates over the $\nu_\mu$ or $\nu_\tau$ flux. For the Gaussian energy resolution function defined in eq.~(\ref{eq:energy_res}), we take $\sigma = 0.11\sqrt{E^{rec}/{\rm MeV}}+0.02\times E^{rec}/{\rm MeV}$, following ref.~\cite{deGouvea:2019goq,Jana:2024lfm}. Unlike in DUNE, Hyper-K will be able to detect incoming $\bar{\nu}_e$ events, using the inverse beta decay~(IBD) process, $\bar{\nu}_e+p\rightarrow e^+ + n$. For Hyper-K, we consider two identical water Cherenkov detectors of 187 kt fiducial mass~~\cite{Hyper-Kamiokande:2018ofw}, equivalent to $N = 6.25\times 10^{33}$ targets. The information on the IBD cross-section is taken from ref.~\cite{Strumia:2003zx}. For energy resolution, we use $\sigma = 0.6\sqrt{E/{\rm MeV}}$. In our calculation, we assume a 20\% overall uncertainty in the neutrino flux~\cite{Jana:2022tsa,Jana:2024lfm}. We also consider statistical uncertainties in our event rate calculation. When the event rate is small ($N<20$), we use Gehrels approximation~\cite{1986ApJ...303..336G} to estimate statistical uncertainties. For the cross-section, we do not consider any uncertainties in this work. Both DUNE and Hyper-K is expected to have a time resolution in nanosecond order~\cite{DUNE:2018jwf,2023NIMPA105568482K}, allowing us neglect the the effect of time resolution.

\begin{figure}
    \centering
    \includegraphics[width=0.99\linewidth]{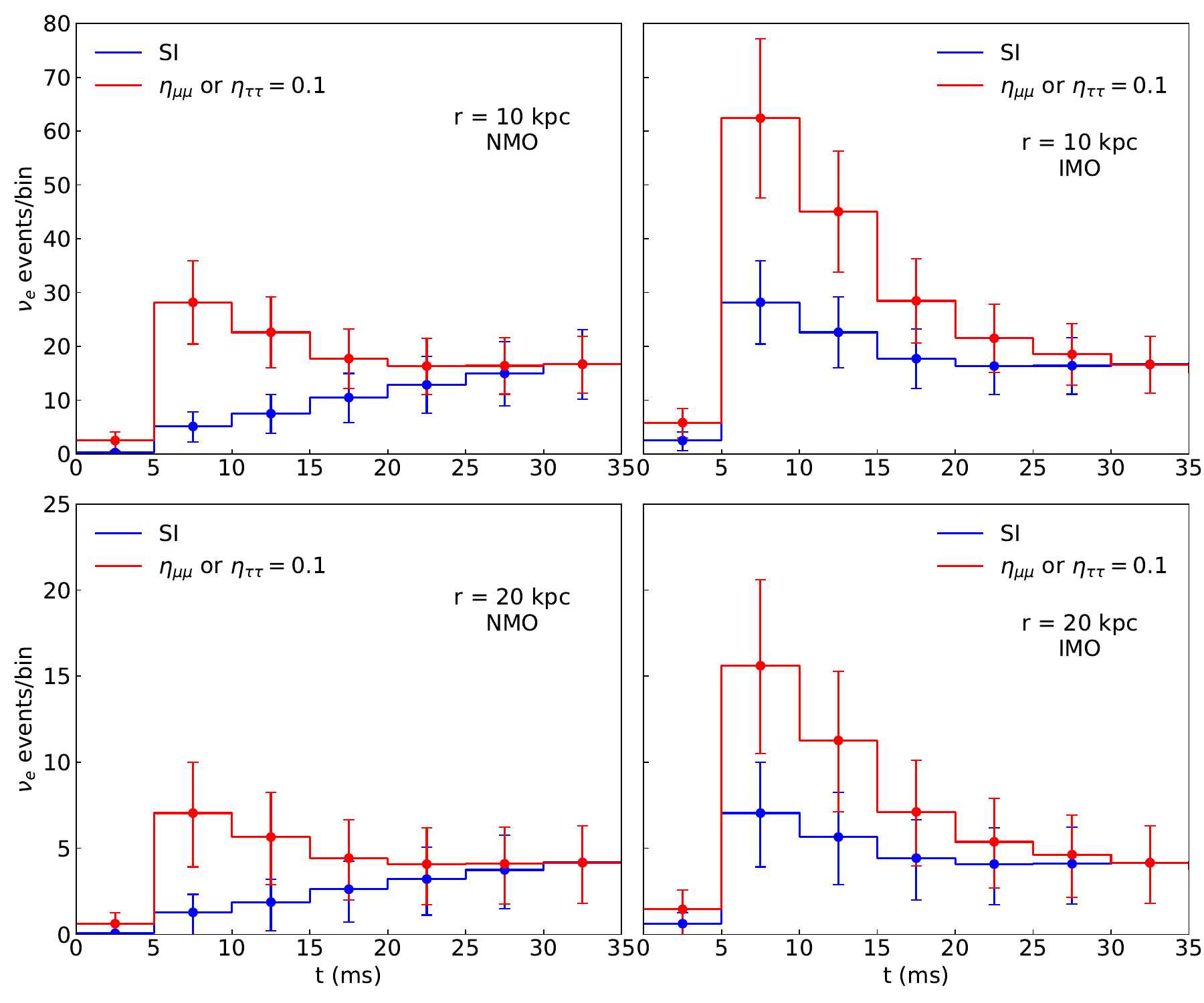}
    \caption{The expected supernova neutrino events~($\nu_e$) distribution at DUNE as a function time after the core bounce. The left~(right) panels shows the distributions for NMO~(IMO). In the top~(bottom) panels, we use 10~kpc~(20 kpc) as distance of the supernova from the Earth.
    For the $\nu_e$ flux, we consider a supernova explosion with $25_\odot$ progenitor mass at a distance 10 kpc from the Earth. The blue line correspond to the case with standard CC interaction~(SI) of neutrinos with background matter in supernova, whereas red line shows the same in the presence of additional SNSI interaction with $\eta_{\mu\mu} = 0.1$ or $\eta_{\tau\tau} = 0.1$. The error bar shows the event rates considering flux and statistical uncertainties. The values of the other oscillation parameters are taken from ref.~\cite{Esteban:2024eli}.}
    \label{fig:event-rate-DUNE}
\end{figure}

In fig.~\ref{fig:event-rate-DUNE}, we show the expected $\nu_e$ event distributions as a function of time after the core bounce at DUNE assuming either NMO~(left panels) or IMO~(right panels). We use two benchmark values of the distance of the supernova from Earth, 10 kpc and 20 kpc as shown in the top and bottom panels.
We plot the distributions for the SI case and also in the presence SNSI parameters $\eta_{\mu\mu}$ or $\eta_{\tau\tau}$ with strength 0.1. 
For the NMO case, $\nu_e \simeq \nu_{3,m}$ in the supernova core in the SI case, leading to an observed flux at Earth proportional to $|U_{e3}|^2$, since the $\nu_e$ flux dominates over the flux of other two flavors. In contrast, in the presence of SNSI, $\nu_e\simeq \nu_{2,m}$ at the supernova core, resulting in a $\nu_e$ flux at Earth proportional to $|U_{e2}|^2$. Since $|U_{e2}|^2\gg|U_{e3}|^2$, we observe a greater number of events in the initial time after the core bounce. At later times, the contributions from the $\nu_\mu$ and $\nu_\tau$ become relevant and reduce the distinction between the two cases. In the IMO case, $\nu_e\simeq \nu_{2,m}$ and $\nu_{e}\simeq\nu_{1,m}$ at the supernova core in the SI case and in the presence of new interaction, respectively. Since $|U_{e1}|^2>|U_{e2}|^2$, we observe more events in the presence of the SNSI interaction than in the SI case. 

Increasing the distance between the supernova and the Earth $R$ by a factor of two reduces the event rate in each time bin by a factor of $1/4$, as shown in the bottom panels, since %It is expected as 
the neutrino flux $f_{\nu_\alpha,\oplus}$ at Earth is proportional to the $1/R^2$. We observe that for $R = 20$ kpc, although the event rate reduces compared to $R=10$ kpc, DUNE can still differentiate between the SI and SNSI scenario in the initial time after the core bounce~($t\lesssim 15$ ms), considering the flux and statistical uncertainties, if the neutrino mass ordering is known. This difference is not prominent anymore for $R> 20$ kpc, as event rates will be extremely small due to the $1/R^2$ suppression. Note that although we show our result for $\eta_{\mu\mu}=0.1$ or $\eta_{\tau\tau} = 0.1$, our results are valid for even smaller value of the SNSI parameters, as long as the new resonance occurs  outside the core of the supernova, which correspond to $\eta_{\mu\mu}\gtrsim 0.006$ or $\eta_{\tau\tau}\gtrsim 0.006$~(see fig.~\ref{fig:Resonance} and related discussion).

\begin{figure}
    \centering
    \includegraphics[width=0.99\linewidth]{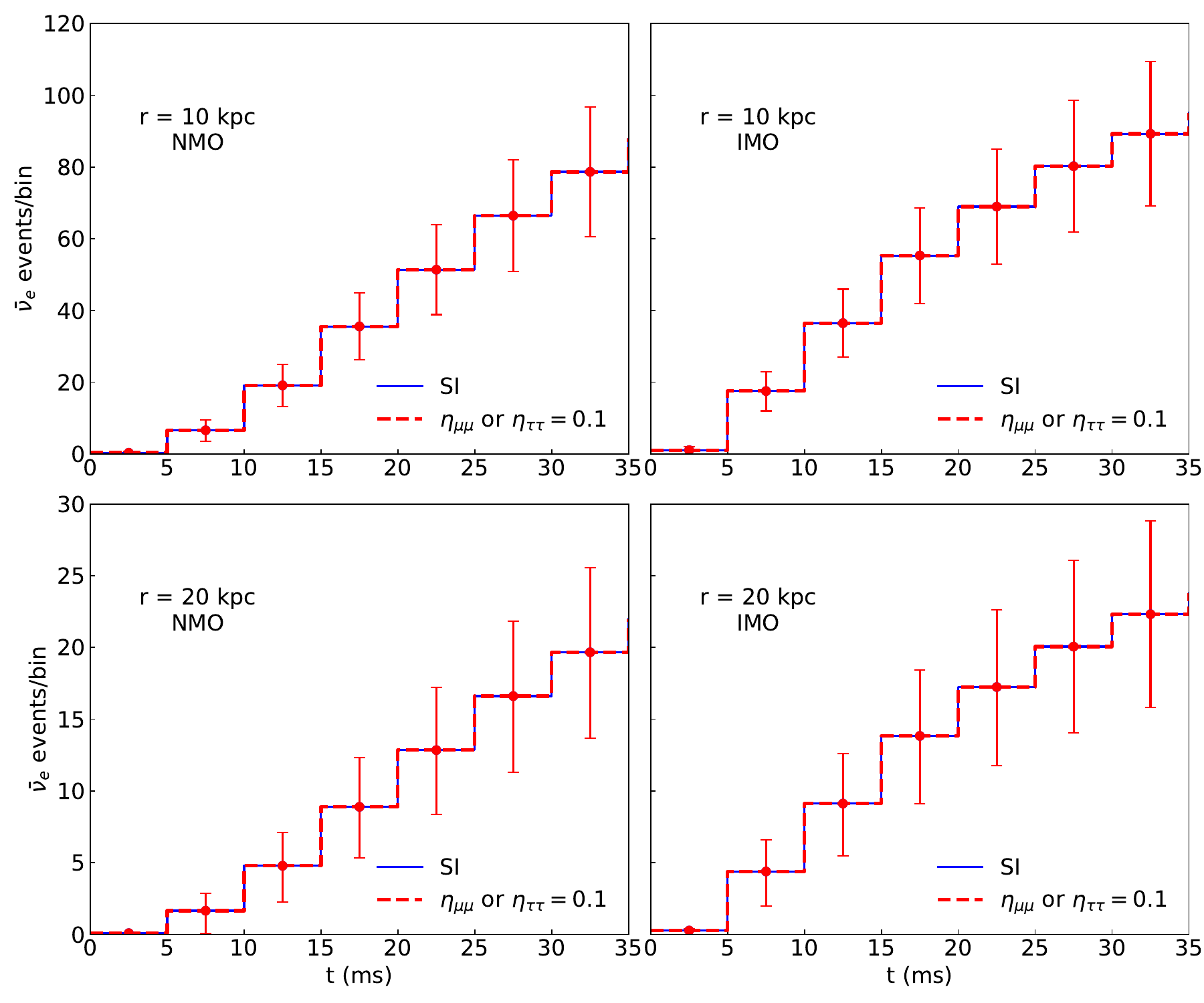}
    \caption{The expected supernova neutrino events~($\bar{\nu}_e$) distribution at Hyper-K with two identical detectors as a function time after the core bounce. The left~(right) panels shows the distributions for NMO~(IMO). In the top~(bottom) panels, we use $R=10$~kpc~(20 kpc) as the distance of the supernova from the Earth. For the $\bar{\nu}_e$ flux, we consider a supernova explosion with $25_\odot$ progenitor mass.  The blue line corresponds to the case with standard CC interaction~(SI) of neutrinos with background matter in supernova, whereas the overlapping red dashed line shows the same in the presence of additional SNSI interaction with $\eta_{\mu\mu} = 0.1$ or $\eta_{\tau\tau} = 0.1$. The error bar shows the event rates considering flux and statistical uncertainties. The values of the other oscillation parameters are taken from ref.~\cite{Esteban:2024eli}.}
    \label{fig:event-rate-Hyper-K}
\end{figure}
One interesting observation is that the event distribution is the same for the SNSI case with NMO and the SI case with IMO. In both cases, the electron neutrino aligns with the same effective mass eigenstate, $\nu_e\simeq\nu_{2,m}$ at the source. Consequently,  the presence of a new interaction ($\eta_{\mu\mu}$ or $\eta_{\tau\tau}$) can brings in a new degeneracy in mass ordering measurement, and DUNE alone can not distinguish the two scenarios. 
The degeneracy that may occur at DUNE can be broken by including complementary observations of $\bar{\nu}_e$ events from a supernova explosion.
 As discussed earlier, the upcoming Hyper-Kamiokande (Hyper-K) experiment will be capable of detecting $\bar{\nu}_e$ via the inverse beta decay process if the $\bar{\nu}_e$ flux is large enough.
In Fig.~\ref{fig:event-rate-Hyper-K}, we show the expected $\bar{\nu}_e$ events in the Hyper-K detector for the $25M_
\odot$ supernova model. For $R = 10$ kpc, we observe a difference of around 10 events in each time bin for the NMO and IMO scenarios in the SI case, which arises because $\bar{\nu}_e$ aligns with $\bar{\nu}_{1,m}$ and $\bar{\nu}_{3,m}$ states, respectively for NMO and IMO. For $R=20$ kpc, the event rate is reduced by a factor of $1/4$ as expected, and the difference between event distribution for two mass ordering scenarios almost vanishes once we consider the flux and statistical uncertainties.
The event distributions remain the same in the presence of the scalar-mediated interaction parameters. It is because, unlike for neutrinos, no new level crossing happens for $\bar{\nu}_e$ flavor evolution in the presence of new physics parameter $\eta_{\mu\mu}$ and $\eta_{\tau\tau}$. However, unlike the DUNE, the  distribution for the SNSI case with NMO is different from the distribution in the SI case with IMO. Therefore, a combined analysis of the event distribution from DUNE and Hyper-K could help distinguish between these two scenarios as long as $R\lesssim 10$ kpc. If the $\nu_e$ event distribution observed at DUNE from a future supernova resembles the distribution shown by the red (blue) curve in the left (right) panel of Fig.~\ref{fig:event-rate-DUNE}, it would infer either NMO with SNSI or the IMO with SI.  However, for $R\lesssim 10$ kpc, a complementary analysis of the $\bar{\nu}_e$ event distribution at Hyper-K may help resolve this degeneracy and identify the true underlying scenario. Other neutrino experiments, like JUNO, that measure $\bar{\nu}_e$ with an energy threshold similar to DUNE or Hyper-K may also help infer the true scenario. 

%==============================
\section{Summary and conclusion}
\label{sec:Summary}
%=============================

This work investigates the flavor evolution of neutrinos in a core-collapse supernova, focusing on the effects of a scalar-mediated, flavor-conserving non-standard interaction between the outgoing neutrinos and the background matter particles alongside the standard model interactions. The new interactions introduce a correction term to the standard Dirac neutrino mass matrix, thereby modifying the neutrino propagation. By studying the evolution of the effective mixing matrix elements, we find that the presence of the SNSI interaction characterized by $\eta_{ee}$ does not have any notable impact on the neutrino flavor evolution, with $\nu_e$ alignment with effective mass state $\nu_{3,m}$ ($\nu_{2,m}$) within the supernova core in the NMO~(IMO) scenario. In contrast, the presence of $\eta_{\mu\mu}$ or $\eta_{\tau\tau}$ leads to a qualitatively different behavior: due to a new level-crossing induced by the interaction, $\nu_e$ aligns with $\nu_{2,m}$ ($\nu_{1,m}$) for normal (inverted) mass ordering as long as $\eta_{\mu\mu}$ or $\eta_{\tau\tau}$ is greater than 0.01. We also find that the evolution of $\bar{\nu}_e$ remains largely unaffected by these new interactions.

We estimate the incoming $\nu_e$ and $\bar{\nu}_e$ fluxes at Earth from a core-collapse supernova originating from a $25M_\odot$ progenitor located at a distance of 10 kpc and 20 kpc. However, for 20 kpc scenario, expected event rate is four times smaller compared to 10 kpc case, as event rate is proportional to $1/R^2$, where $R$ is the distance between the Earth and the supernova.
We compute the estimated event distribution as a function of the time after the core bounce at the next-generation neutrino experiments DUNE and Hyper-K. At DUNE, $\nu_e$ events are primarily detected through charged-current (CC) interactions with argon nuclei, while Hyper-K is sensitive to $\bar{\nu}_e$ via inverse beta decay. Our results show that in DUNE, the expected event rate for neutrinos is significantly higher in the presence of SNSI parameter $\eta_{\mu\mu}$ or $\eta_{\tau\tau}$ during the initial time after the core bounce, for both mass orderings of neutrinos. 
Considering 20\% flux uncertainties and statistical uncertainties, we find that DUNE would be able distinguish between the SI and SNSI scenario if the mass ordering is known. However, a new degeneracy in the event distribution arises in the presence of a new scalar-mediated interaction ---  the event distribution for normal mass ordering with SNSI matches that of the inverted mass ordering with standard interactions. We demonstrate that $\bar{\nu}_e$ events observed at the Hyper-K detector may help to lift the degeneracy if the supernova distance is small and neutrino luminosity is large enough. In the Hyper-K detector, we observe that for the IMO scenario, the expected event rates in each time bin are larger than those in the NMO scenario. However, the SNSI parameters do not have any impact on the $\bar{\nu}_e$ event distribution. Therefore, a combined analysis can help distinguish between the NMO+SNSI and the IMO+SI scenarios. 

The observation of neutrino events from Supernova 1987A has been a groundbreaking achievement in the field of multi-messenger astrophyics, offering crucial insights into the dynamics of supernova explosion as well as neutrino properties. Next-generation neutrino detectors, such as DUNE and Hyper-K, with their enhanced sensitivity and detection capabilities, are expected to observe a large number of neutrino events, improving our understanding of neutrino astrophysics and probing physics beyond the standard model.

%=============================%
\subsubsection*{Acknowledgments}
%=============================%

We would like to thank Sudip Jana, Yago Porto, Bhaskar Dutta, Aparajitha Karthikeyan, and Nityasa Mishra for discussions. We also thank Damiano F. G. Fiorillo and Edoardo Vitagliano for insightful comments. S.D. thanks the organizers of the PHENO-2025 conference at the University of Pittsburg for providing an opportunity to present the preliminary results of this work. This work is supported by the U.S. Department of Energy Grant DE-SC-0010113.

%===============================
\begin{appendix}
%\label{appendix:adiabaticity}
%=============================%

\section{Adiabatic propagation neutrinos in the presence of scalar NSI}
\label{app:A1}
\noindent
 For adiabatic neutrino propagation through a non-uniform medium like supernova, adiabaticity condition  must be satisfied at the resonance. In this appendix, we test the adiabaticity condition for neutrino propagating from the supernova core to the surface in the presence of scalar NSI. Considering ($e,\mu$) block of a three-flavor neutrino Hamiltonian, the adiabaticity condition is given by~\cite{Jana:2024lfm} 
 \begin{align}
     \gamma = \left|\frac{4 |\mathcal{H}_{e\mu}|^2}{\dot{\mathcal{H}}_{ee}-\dot{\mathcal{H}}_{\mu\mu}}\right|>1\,,
     \label{eq:adiabaticity}
 \end{align}
 where $\mathcal{H}_{\alpha\beta}$ denotes the elements of the Hamiltonian defined in eq.~(\ref{eq:Hs}), which we further expand,
 \begin{align}
     \mathcal{H} = \frac{1}{2E}\left[MM^\dagger + \delta M M^\dagger +\delta M^\dagger M + \delta M \delta M^\dagger+2E\cdot V_{\rm SI}\right]\,,
     \label{eq: Hamiltonian}
 \end{align}
where $M$, $\delta M$, and $V_{\rm SI}$ are defined in section~\ref{sec:SNSI-formalism}. The first term in eq.~(\ref{eq: Hamiltonian}) shows the vacuum part of Hamiltionian, whereas the other three terms are comprised of the standard matter interaction and the SNSI. In the inner region of the supernova, where the matter density is large, the matter interaction part of the Hamiltonian dominates. Considering only the SNSI parameter $\eta_{\mu\mu}$ to be non-zero, and neglecting the vacuum terms in the Hamiltonian, we get
 \begin{align}
     \mathcal{H}_{ee} &\simeq V_{\rm CC}=\sqrt{2}G_F N_e= \sqrt{2}G_F\frac{\rho}{m_n}Y_e \,,\\
     \mathcal{H}_{e\mu}  &\simeq \frac{1}{2E} M_{e\mu} \delta M_{\mu\mu} = \frac{1}{2E} M_{e\mu} \delta\tilde{M}_{\mu\mu}\frac{\rho}{\rho_s}\,,\\
     \mathcal{H}_{\mu\mu} &\simeq \frac{1}{2E}\delta M_{\mu\mu}\delta M_{\mu\mu}= \frac{1}{2E} \left[\delta \tilde{M}^2_{\mu\mu}\left(\frac{\rho}{\rho_s}\right)^2\right]\,.
\end{align}

$M_{\alpha\beta}$~($\delta\tilde{M}_{\alpha\beta}$) denotes elements of the matrix $M$~($\delta\tilde{M}$), defined in  eq.~\ref{eq:scaling_density}. Note that in the expression for $\mathcal{H}_{\mu\mu}$, we 
neglect the term proportional to $\delta M$, which is small compared to the term with $\delta M_{\mu\mu} \delta M_{\mu\mu}$ at large densities. It is clear that the denominator in eq.~(\ref{eq:adiabaticity}) is proportional to $d\rho/d r$, considering $Y_e$ to be constant.
Following the density profile in ref.~\cite{Serpico:2011ir,Jana:2024lfm}, which approximately follows $\rho(r) \approx \rho_c \left(\frac{r}{R_c}\right)^{-3}$, where $\rho_c = 10^{14}$ g/cc and $R_c = 10$ km, we estimate $\gamma$ in eq.~(\ref{eq:adiabaticity}). After simplification, approximate expression for the $\gamma$ is,
\begin{align}
    \gamma \simeq \frac{6.25\times 10^{-6}(\frac{r}{R_c})^{-6}}{\left|-3.2\times10^{-10}\left(\frac{r}{R_c}\right)^{-4}+1.5\times 10^{-4}\left(\frac{r}{R_c}\right)^{-7}\right|}\,.
\end{align}

We check that adiabaticity condition is satisfied at the density corresponding to the new resonance in the presence of SNSI parameters, ($\rho \approx 10^8~(10^{10})$  g/cc for $\eta_{\mu\mu} = 0.1~(0.01)$) So it is safe to assume adiabatic propagation in our work.

%$r~\gtrsim 60$ km ($\rho \lesssim 10^{12}$ g/cc), which is well above density at with density at which new transition between the $U^{\rm eff}_{e2}$ and $U^{\rm eff}_{e3}$ occurs in the presence of $\eta_{\mu\mu}$ or $\eta_{\tau\tau}$ with strength 0.1. So it is safe to assume adiabatic propagation in our work.
\end{appendix}

\bibliographystyle{JHEP}
\bibliography{ref}

\end{document}